# Ex Vivo Mouse Brain Microscopy at 15T with Loop-Gap RF Coil

*Ouri Cohen[1,2] and Jerome L. Ackerman[1,2]*

[1]Athinoula A. Martinos Center for Biomedical Imaging
Department of Imaging
Massachusetts General Hospital
149 13th Street, Charlestown, MA 02129 USA
[2]Department of Radiology, Harvard Medical School, Boston, MA 02115 USA

**ABSTRACT**

The design of a loop-gap-resonator RF coil optimized for ex vivo mouse brain microscopy at ultra high fields is described and its properties characterized using simulations, phantoms and experimental scans of mouse brains fixed in 10% formalin containing 4 mM Magnevist™. The RF ($B_1$) and magnetic field ($B_0$) homogeneities are experimentally quantified and compared to electromagnetic simulations of the coil. The coil's performance is also compared to a similarly sized surface coil and found to yield double the sensitivity. A three-dimensional gradient-echo (GRE) sequence is used to acquire high resolution mouse brain scans at 47 $\mu m^3$ resolution in 1.8 hours and a 20×20×19 $\mu m^3$ resolution in 27 hours. The high resolution obtained permitted clear visualization and identification of multiple structures in the ex vivo mouse brain and represents, to our knowledge, the highest resolution ever achieved for a whole mouse brain. Importantly, the coil design is simple and easy to construct.

# 1.Introduction

Magnetic resonance microscopy (MRM) is a tool for microscopic imaging of ex vivo soft tissue samples. Unlike histology, MRM allows imaging of multiple tissue contrasts ($T_1$, $T_2$, diffusion-weighted, etc.) in addition to morphometric measurements. Moreover, because it is non-invasive, MRM preserves structural spatial relationships that can be lost during the sectioning process used in histology. Nevertheless, achieving micron-scale resolutions in reasonable scan times remains a technical challenge given the intrinsically poor signal to noise ratio (SNR) of MRI and the rapid degradation of SNR with increasing magnification. Overcoming this limitation requires the use of strong magnetic fields and special consideration in the design of the RF coil. One possible way of mitigating the low SNR intrinsic to smaller voxels is through reduction of the coil size [1]. This has led to development of micro-coils used to obtain in-plane resolutions as small as 3 µm [2]. However, the increased signal per pixel comes at the cost of reduced spatial coverage that prevents the use of micro-coils for experiments with large field-of-views (FOV). Instead, volume coils must be used with the added benefit of homogenous RF transmission and reception. Solenoid coils in particular are a popular choice and have been used with different configurations and field strengths in the quest to reduce the achievable voxel size. Huang et al., [3] reported scan times of 11 hr imaging fixed mouse brains at 14T, achieving $25 \times 25 \times 37$ µm in a $512 \times 512 \times 256$ matrix, and $25 \times 25 \times 37$ µm in a $400 \times 256 \times 256$ matrix. An in-plane resolution of 39 µm was obtained in an ex vivo mouse brain in a 10 hour scan at 9.4 T using a custom made solenoid coil [4]. More recently, 22 µm resolution was achieved, with the long scan time offset by partial acquisition of Fourier space though at the cost of some blurring [5]. Similar resolutions were obtained at a lower, 7 T, field strength with a solenoid coil but with a longer (13 hour) scan time [6].

In this work we propose an alternative to the solenoid coil and describe the construction of a remotely-tunable bridged double loop-gap resonator (LGR) RF coil optimized for fixed mouse brains at 15 T. The loop-gap resonator design was chosen for its simple construction, large frequency tuning range, good filling factor and excellent RF homogeneity [7–10]. In particular, this double LGR design is based on that reported by Koskinen and Metz [8]. We compare the sensitivity of our design to that of a similarly sized surface coil. The combination of high field strength and optimized RF coil allowed acquisition of $20 \times 20 \times 19$ μm$^3$ voxels that, to our knowledge, represent the highest resolution yet achieved in whole mouse brain MR imaging.

## 2. Material and Methods

### 2.1 Coil Construction

Coaxial cylinders made of plastic (outer cylinder) and quartz (inner cylinder), with dimensions as shown in Figure 1a, were used as formers for the coupled resonator devices. The cylindrical conductors and bridge were formed from self-adhesive copper foil tape (Scotch Brand Foil Tapes, 3M, Inc., Minneapolis, MN, USA) with gaps of 2 and 5 mm in the inner and outer cylinders respectively. The gap widths were chosen to adjust the resonance frequency of the resonator to that of the 619 MHz scanner. Adhesive Teflon tape was used to secure the copper tape, allow smooth relative rotation of the two cylinders and to increase the dielectric constant of the inter-cylinder space. The gap in the inner cylinder was bridged by a 24 mm long × 8 mm wide copper tape to improve the magnetic field homogeneity [9]. Rotating the resonators with respect to each other allowed gross tuning of the resonance frequency of the coil over a 400 MHz range.

Since the small diameter of the magnet bore permitted limited access to the coil when inside the magnet, a gear and rack system was used to adjust the distance of a separate flat gapped tuning loop from the end of the resonator to achieve fine tuning of the coil [7]. Coupling to the scanner for transmission and reception (TR/RX) was accomplished with a coaxial loop placed at the opposite ends of the resonators from the fine tuning loop, and connected to a 50 ohm coaxial cable. Adjusting the separation of the coupling loop from the resonator achieves impedance matching. In an alternative design, impedance matching may be achieved by adjusting a variable capacitance in series with a fixed coupling loop. In both coupling alternatives, the use of inductive coupling between the scanner and the resonator makes the resonator and specimen electrically balanced with respect to ground. This reduces the effect of parasitic capacitance of the resonator and specimen to ground, and further reduces the tendency of the resonator to radiate. Finally, the inherent symmetry of this type of coupling improves the $B_1$ uniformity.

### 2.2 Simulation

A model of the coil was created in HFSS (Ansys, Inc., Canonsburg, PA, USA) and the magnitude of the $B_1$ field calculated for a homogenous 13 mm diameter × 24 mm high cylindrical distilled water phantom. Each $B_1$ field map was normalized to its maximum value set to unity to facilitate comparison. The mean and standard deviation of the normalized field were computed to obtain a measure of the RF homogeneity.

**2.3 Phantom**

Experiments were conducted on a 15 T 130 mm horizontal bore Magnex Ltd. (now Agilent, Yarnton, Oxford, UK) magnet equipped with a Resonance Research, Inc., (Billerica, MA, USA) 60 mm ID gradient insert with 2370 mT/m maximum gradient, interfaced to a Siemens Medical Systems (Erlangen, Germany) console.

A 13 mm diameter $\times$ 35 mm long water-filled NMR tube (Wilmad-LabGlass, Inc., Vineland, NJ, USA) was used as a phantom for characterizing the $B_0$ and $B_1$ maps of the coil. Gradient recalled echo (GRE) images at two echo times TE=10 and 10.49 ms were used in the Siemens field mapping sequence to compute the $B_0$ field map according to the formula provided in [11]. The manufacturer's built in mapping sequence was used to compute the $B_1$ map. The homogeneities of the experimental $B_0$ and $B_1$ maps were measured in the central plane by calculating the mean and standard deviation, same as for the simulated maps as described above. The FOV for both experiments was set to $25 \times 25$ mm with a matrix size of $128 \times 128$ giving a voxel size of 195 $\mu m^2$.

**2.4 Mouse Brain**

*2.4.1 Specimen Preparation*

Freshly dissected mouse brains were purchased from ThermoFisher Scientific, Inc., (Pittsburgh, PA, USA), placed in Fisher sodium acetate buffered 10% formalin for about one week, and then transferred to a 4 mM solution of Magnevist (gadolinium DTPA, Bayer HealthCare Pharmaceuticals, Berlin, Germany) in the 10% buffered formalin solution for an additional week to shorten relaxation times and improve contrast. The brains were patted dry and transferred to 13 mm glass NMR tubes containing a proton-free liquid fluorocarbon (Fomblin perfluoropolyether, Ausimont, Thorofare, NJ, USA) to eliminate the intense proton signal from the fixation medium, mitigate susceptibility-difference artifacts at tissue-air interfaces, and avoid specimen dehydration. The sample was then degassed in a 50 Torr vacuum to remove trapped air bubbles in the tissue and further reduce susceptibility artifacts.

*2.4.2 Data Acquisition*

Two separate experiments were conducted. In the first experiment a 3D GRE sequence was used with the following parameters: flip angle α=52°, TE = 3.45 ms, NEX = 4, BW=260 Hz/pixel. The FOV was set to 12 mm$^3$ with a matrix size of 256×256×256 yielding a voxel size of $47 \times 47 \times 47$ μm$^3$ (voxel volume = 103.8 pL). The TR was set to 25 ms yielding a total scan time of 1.8 h.

The same GRE sequence was used for the second experiment but with the following parameters: flip angle α=45°, TE = 3.79 ms, NEX = 8, BW=260 Hz/pixel. The FOV was set to 12.4 mm$^3$ with a matrix size of 620×620×640 yielding a voxel size of $20 \times 20 \times 19$ μm$^3$ (voxel volume = 7.6 pL). The TR was set to 30 ms which was the shortest achievable on this system without overheating the gradient coil. The relatively long TR resulted in a total scan time of 27 h.

*2.4.3 Drift Compensation*

Ultrahigh field magnets often exhibit significant field drift which is typically corrected with the deuterium field-frequency lock or field ramp setting of a standard spectrometer console. The Siemens console, designed for much lower field magnets and much shorter scan times, has no drift compensation capability. To correct for the high drift rate of the 15T magnet we implemented a linear drift compensation which digitally ramps the local oscillator (reference) frequency of the homebuilt transceiver to match the magnet's natural decay of about 100 Hz/hr, reducing the effect of magnet drift to an insignificant level over a 24 hour acquisition.

*2.4.4 Post-Processing*

The acquired GRE images were rigidly rotated and translated in post-processing using the Multi-image Analysis GUI software package (University of Texas Health Science Center, San Antonio, TX, USA) to correct for variable specimen placement.

**2.5 Comparison with surface coil**

The sensitivity of our coil was compared to that of a 13.6 mm inner diameter single loop, flat surface coil. By the principle of reciprocity [12] the relative sensitivity of the two coils can be calculated as the inverse ratio of the voltage required to achieve a 90° excitation pulse for each coil. Since the sensitivity of a loop depends on its diameter, the relative sensitivity was scaled by the ratio of the diameters of the coils for purposes of comparison.

A coronal slice was acquired with each coil using a GRE sequence and identical acquisition parameters (flip angle $\alpha=82°$, TE = 3.4 ms, NEX = 1, FOV= 15 mm$^2$, matrix size = 512×512 and TR=200 ms, BW=255 Hz/pixel). To illustrate the impact of the different sensitivity of the coils, the ratio of SNRs of the two images was also calculated.

# 3. Results

**3.1 Coil Characterization**

Axial slices from the experimental and simulated $B_1$ maps are shown in Figure 2a-b. The mean and standard deviation of the normalized experimental $B_1$ map was $0.91 \pm 0.18$, similar to the $0.93 \pm 0.05$ of the normalized simulated map. The calculated $B_0$ map (Figure 3) had a mean and standard deviation of $0.03 \pm 0.14$ ppm.

### 3.2 Comparison with surface coil

The ratio of voltages required to achieve a 90° pulse for the two coils was $V_{LGR}/V_{Surface}$ = 3.7V/7.9V, implying a 2.1-fold sensitivity gain for the LGR coil. However, given the smaller diameter of the surface coil loop, the actual gain is scaled by $D_{LGR}/D_{Surface}$ = 14mm/13.6mm = 1.03, resulting in a total sensitivity gain of 2.2. Mouse brain images acquired with each coil are shown in Figure 3a-b. Despite the identical acquisition parameters, the image acquired using the surface coil is noisier with an SNR ratio for the two coils of $SNR_{LGR}/SNR_{Surface}$ = 2.0.

### 3.3 Mouse Brain

The acquired mouse brain data is shown in Figure 4 for the two image resolutions tested. A comparable slice from the Allen Mouse Brain Atlas [13] is shown as well for comparison and identification of the various brain structures visible in the images. Some of the visible structures include the corpus callosum, dentate gyrus and the stratum lacunosum-moleculare.

## 4. Discussion

We have demonstrated an LGR based RF coil design optimized for high resolution scanning of mouse brain specimens at ultra high field. The coil benefits from both an increased sensitivity relative to a surface coil, as well as excellent $B_1$ homogeneity, as shown by the simulated and experimental $B_1$ maps (Figure 2a-b). Coil dimensions were expressly selected to maximize the filling factor of the mouse brain specimens which yielded the highest resolution images (Figure 4) yet obtained for whole brain MR mouse imaging.

Although it was first introduced nearly thirty years ago, to our knowledge this is the first application of an LGR based coil design at 15 T. Imaging at 15 T is challenging in part because of huge magnetic susceptibility effects. To obtain high magnifications at useful SNR, coil performance, and in particular its tuning, must be optimized. Because they contain heavy metals with large magnetic susceptibilities, conventional discrete variable and fixed tuning capacitors, as well as solder joints, can contribute to shimming problems and can create susceptibility dropouts. Conventional solenoid coils of more than one or two turns around centimeter-sized specimens may be close to self resonance at high frequencies, in which case they will perform poorly. Additionally, the electric fields of conventional solenoid coils will penetrate the specimens, further degrading performance. The double concentric loop-gap resonator is effectively self-shielded electrostatically, and minimizes the penetration of electric fields into the specimen (minimizing specimen heating and the Johnson noise associated with the resistive component of the coil impedance). Separate discrete tuning capacitors are not required, and the electric fields of the circuit capacitance are confined to a region between the cylinders and away from the specimen. Additionally, the double concentric loop-gap resonator is effectively a one-turn solenoid; minimizing the inductance by minimizing the turn count of a solenoid often maximizes its performance at high fields in that it operates well below self resonance (is most purely inductive). Frequency tuning of the resonator is accomplished by relative rotation of the two cylinders. Impedance matching of the resonator is accomplished by an innovative gear and rack mechanism which moves a separate gapped flat loop axially with respect to the resonator, via a knob conveniently located at the end of the probe close to the magnet opening.

Given its ability to non-destructively measure different contrasts and morphologies and combined with advances in genetic manipulation methods, interest in mouse brain microscopy is increasing. This is particularly true because many of the techniques used can be translated into clinical practice. Indeed, MRI has been used with transgenic mouse models of Miller-Dieker syndrome [14], Alzheimer's [15,16] and brain edema [17]. The high resolution and simple construction of our coil allows precise measurements of distinct brain regions which may help our understanding of changes resulting from different genetic modifications.

This coil design suffers from some limitations. The coil dimensions are optimized for mouse brain specimens and are therefore not suitable for larger samples or in vivo imaging. Although coil dimensions can be modified, increasing the coil size would result in a reduction in the filling factor for some samples. Nevertheless, given the high SNR obtained with this coil, we have successfully scanned samples as small as 3 mm in diameter [18]. Further gains in SNR can be obtained by placing a preamplifier closer to the coil to avoid the losses in the cables given the high operating frequency, and is the subject of future work. Finally, the relatively long scan time is a result of limitations in the gradient cooling capability of our system. To reduce the gradient power dissipation, TR is set longer than the value dictated by the other scan parameters, and could be readily reduced by a factor of 2-3 with improved cooling of the gradient coil.

## 5. Conclusion

We have described and demonstrated the design of an RF coil optimized for scanning mouse brain specimens. The high SNR afforded by the coil allowed us to obtain images at a resolution of $20 \times 20 \times 19$ μm$^3$, to our knowledge the highest resolution achieved at 15 T thus far. The coil design is simple and easy to construct and allows remote tuning even when inside the magnet.

## 6. Acknowledgments

This research was supported in part by the Athinoula A. Martinos Center for Biomedical Imaging and the Center for Functional Neuroimaging Technologies (funded by NIH grant P41-EB015896). The 15T scanner was developed with High End Instrumentation Grant S10RR023009 from the National Center for Research Resources of the NIH. OC gratefully acknowledges support from USAMRMC Defense Medical Research and Development Program (DMRDP) award W81XWH-11-2-0076 (DM09094) and NIH grant R44-AR065903. We are grateful for specimens prepared by Dr. Frederick A. Schroeder.

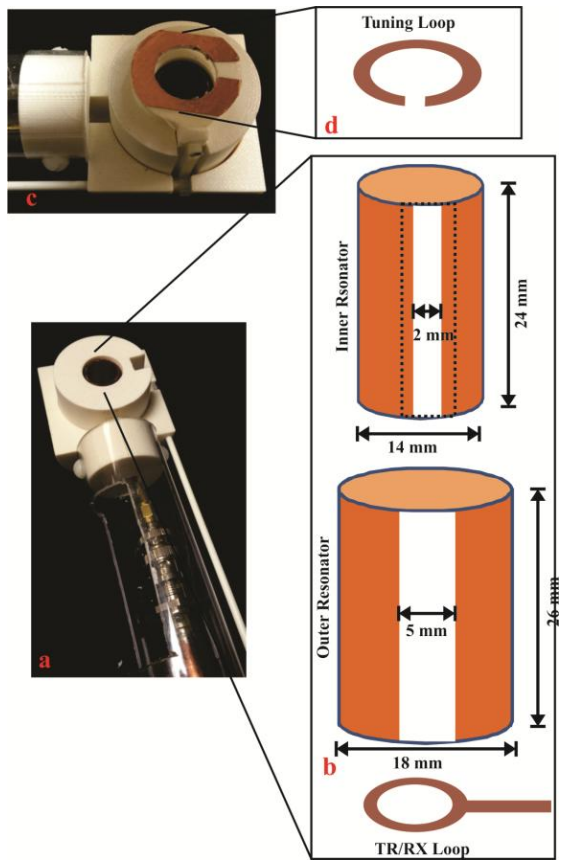

**Fig. 1**. Photographs of the double LGR RF coil design (a), (c) along with exploded view illustrations (b),(d). Relative rotation of the concentric resonators provides gross tuning of the frequency of the coil by varying the overlap across the capacitive gaps in the copper foil covering the resonators, in effect varying the capacitance in series with the coil inductance. Note that no discrete capacitors are needed for tuning. An additional piece of copper foil (dotted rectangle), insulated from both cylinders and rotating with the inner cylinder, was used to bridge the gap on the inner resonator and improve the RF homogeneity. Adjusting the height of the loop above the coil with the gear and rack mechanism (c) allowed fine tuning of the resonance frequency of the coil while inside the scanner. Coupling to the scanner was achieved using the TR/RX loop shown.

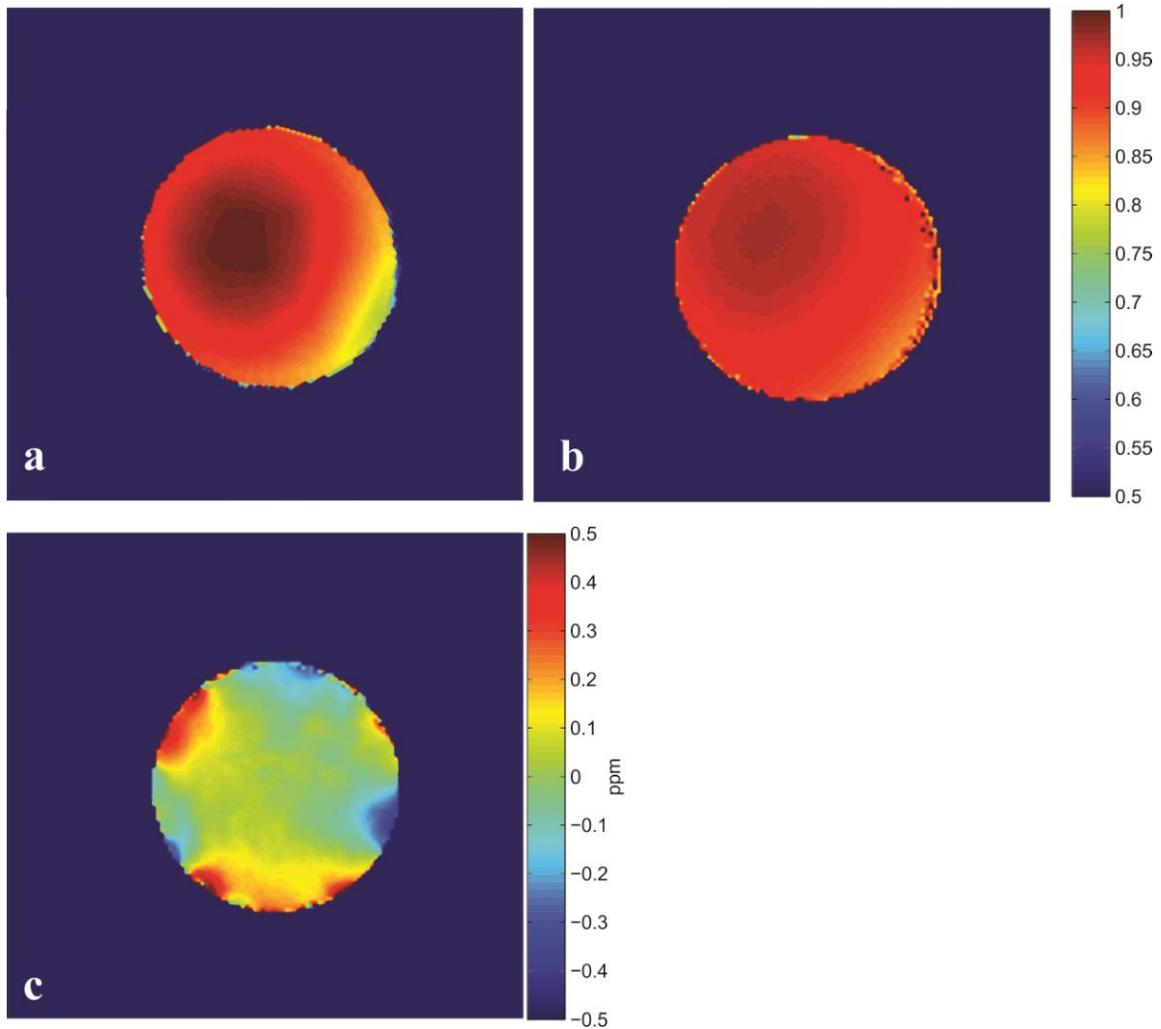

**Fig. 2.** Normalized simulated (a) and experimental (b) $B_1$ maps, and experimental $B_0$ map in ppm (c) obtained using a 13 mm diameter distilled water phantom. The means and standard deviations of the simulated and experimental $B_1$ maps were $0.93 \pm 0.05$ and $0.91 \pm 0.18$ respectively; the maximum in each map was set to unity. The mean and standard deviation of the $B_0$ field was $0.03 \pm 0.14$ ppm.

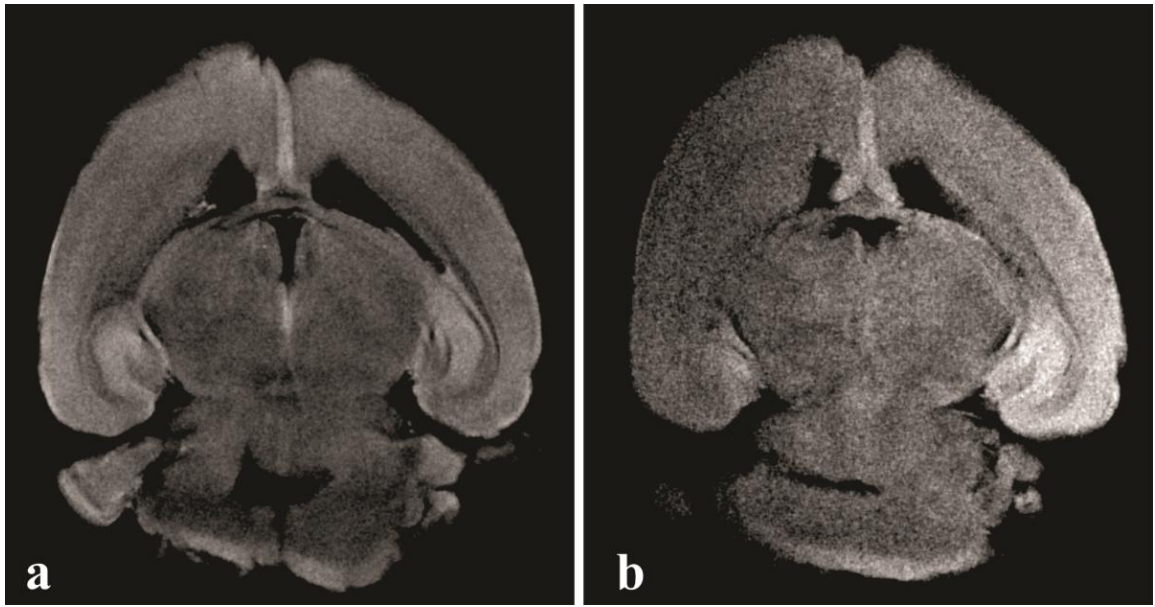

**Fig. 3.** Coronal slices from the mouse brain acquired with the LGR (a) and surface (b) coils using the following acquisition parameters: flip angle α=82°, TE = 3.4 ms, NEX = 1, FOV= 15 mm$^2$, matrix size = 512×512 and TR=200 ms. Despite the use of identical acquisition parameters the LGR coil yielded 2× higher SNR .

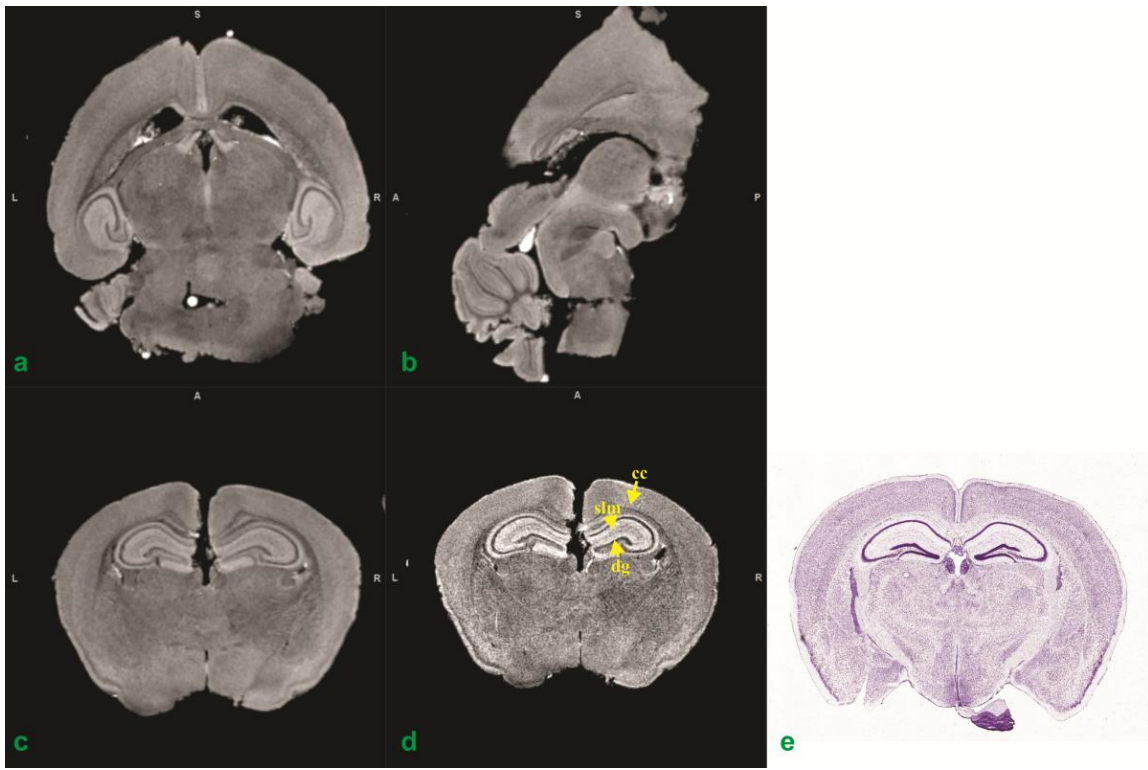

**Fig. 4.** Coronal (a), Sagittal (b) and Axial (c) slices of the mouse brain data acquired with a 47 μm$^3$ resolution in 1.8 hours. A higher resolution (20 μm$^2$) axial slice from the longer (27 h) acquisition is shown in (d). A comparable Nissl stained slice from the Allen Mouse Brain atlas is shown in (e) for comparison. Note the fine details revealed by this high spatial resolution with a number of brain structures, denoted by yellow arrows, visible which include the corpus callosum (cc), dentate gyrus (dg) and the stratum lacunosum-moleculare (slm).